\documentclass[conference]{IEEEtran}
\usepackage{algorithm} 
\usepackage{algpseudocode}
\usepackage{graphicx}
\usepackage{amssymb}
\usepackage{amsfonts}
\usepackage{amsmath}
\usepackage{float}
\usepackage{bm}
\usepackage{nohyperref}

\IEEEoverridecommandlockouts

\newtheorem{Example}{Example}

\ifCLASSINFOpdf
\else
\fi
\begin{document}
	\title{A Neural Network Lattice Decoding Algorithm}
	\author{Mohammad-Reza~Sadeghi, Farzane Amirzade, Daniel Panario, and Amin Sakzad \\
		\thanks{M.-R.~Sadeghi and F.~Amirzade are with the Faculty of Mathematics and Computer Science, Amirkabir University of Technology, Tehran, Iran. D.~Panario is with the School of Mathematics and Statistics, Carleton University, Ottawa, Canada. A.~Sakzad is with the Clayton School of IT, Monash University, Melbourne, Australia.
			(e-mails:  msadeghi@aut.ac.ir, famirzade@gmail.com, daniel@math.carleton.ca, amin.sakzad@monash.edu).}
	}

	
	\maketitle
	\pagestyle{empty}
	\thispagestyle{empty}
	\begin{abstract}
		Neural network decoding algorithms are recently introduced by Nachmani et al.
		 to decode high-density parity-check (HDPC) codes. In contrast with iterative decoding algorithms such as sum-product or min-sum algorithms in which the weight of each edge is set to $1$, in the neural network decoding algorithms, the weight of every edge depends on its impact in the transmitted codeword. In this paper, we provide a novel \emph{feed-forward neural network lattice decoding algorithm} suitable to decode lattices constructed based on Construction A, whose underlying codes have HDPC matrices. We first establish the concept of feed-forward neural network for HDPC codes and improve their decoding algorithms compared to Nachmani et al. We then apply our proposed decoder for a Construction A lattice with HDPC underlying code, for which the well-known iterative decoding algorithms show poor performances. The main advantage of our proposed algorithm is that instead of assigning and training weights for all edges, which turns out to be time-consuming especially for high-density parity-check matrices, we concentrate on edges which are present in most of $4$-cycles and removing them gives a girth-$6$ Tanner graph. This approach, by slight modifications using updated LLRs instead of  initial ones, simultaneously   accelerates the training process and improves the error performance of our proposed decoding algorithm.
	\end{abstract}

	%
	\IEEEpeerreviewmaketitle
	\section{Introduction}
	
	\IEEEPARstart{C}onstructing lattices from codes has been an intense research topic which resulted in well-know Constructions A, B, C, D, and D' (see \cite{Barnes} and \cite{Conway}). High dimensional lattices along with iterative decoding algorithms were first introduced by Sadeghi et al. \cite{Sadeghi}. Although high dimensional lattices have high coding gain, in practice, they are difficult to implement and in higher dimensions they have high decoding complexity. 
	Several decoding algorithms have been proposed in the literature. The generalized min-sum algorithm was presented in \cite{Sadeghi} to decode low-density parity-check (LDPC) lattices. Other iterative decoding algorithms to decode LDPC lattices are sum-product algorithm (SPA) \cite{Korea} and FFT based SPA \cite{Lida}. In these methods, the process of transmitting messages in variable node and check node operations is applied on the Tanner graph  of LDPC lattices.
	
	Some underlying codes of the algebraic lattices have a rather high-density parity-check (HDPC) matrix. For example, the underlying code of Barnes-Wall lattices are Reed-Muller codes whose parity-check matrices are not sparse. These lattices are based on Construction D' \cite{Forney} using a set of nested Reed-Muller codes. In this case, the variable-node and check-node operations in message passing decoding algorithms are time-consuming. In fact, applying iterative algorithms to decode most well-known algebraic codes would result in poor performance when compared to maximum likelihood decoders~\cite{ViB}. To resolve this problem, deep neural network decoders were proposed~\cite{Learning2016,Learning1}.

	In recent years, deep learning methods have shown amazing performances in a variety of subjects. For example, an application of deep learning methods to the problem of low complexity channel decoding has been proposed \cite{Learning1}. It has been demonstrated that deep learning methods improve the min-sum and sum-product algorithms for HDPC codes using a weighted trellis graph,  which is another representation of codes. The main benefit of the neural network decoder is that the weight of every edge relies on its influence in the transmitting messages. By setting weights properly, we can compensate for small cycles, which are the main causes of high error floor regions and deterioration of the decoding process. The first step in this method is to train weights for a given codeword like the all-zero codeword ${\bf 0}$. Then, using these trained weights, we decode codewords. The computation complexity of the neural network decoder depends on the number of weights required to be trained. For example in \cite{Learning2016}, this number is $2n+2LE$, where $n$ is the number of variable nodes, $L$ is the number of iterations and $E$ is the number of edges in the corresponding trellis diagram of the code. 
	
	In the $\ell$-th iteration of trellis graph in \cite{Learning2016}, there are two hidden layers $VN_\ell$ and $CN_\ell$ for the variable node operation and the check node operation, respectively. In this framework, two weights $w_\ell$ and $w'_\ell$ are defined for each edge in the $\ell$-th iteration. The weight $w_\ell$ links the first layer and the hidden layer, $CN_\ell$, and the weight $w'_\ell$ links the hidden layers $CN_\ell$, $VN_{(\ell+1)}$. In \cite{Learning1}, a new recurrent neural network (RNN) algorithm was defined by which the number of weights to train was significantly reduced to $2E$. In this method, all weights between the first layer and hidden layers $CN_\ell$, for $\ 1\leq \ell\leq L$, are set to $1$. No training process is needed for them as the experimental results have shown that training these parameters did not result in any performance improvement. Moreover, for each edge in the RNN method, two weights $w$ and $w'$ are considered fixed in iterations. In this framework, $w$ is the weight associated to the hidden layers $CN_\ell$, $VN_{(\ell+1)}$ and $w'$ is the weight associated to the hidden layer $CN_\ell$ and the layer related to hard decision operation of the $\ell$-th iteration. Numerical results \cite{Learning1} show that the RNN method outperforms the neural network decoding algorithm in \cite{Learning2016}.
	 
	In a concurrent work, a deep learning lattice decoder is also proposed in \cite{NLD}. In this paper, we aim to present a neural network decoding algorithm for lattices constructed based on Construction A, whose underlying codes have high-density parity-check matrices. All the weights $w'$ in the RNN algorithm are trained in this work. However, instead of considering the weight $w$ for each edge, similar to the RNN algorithm, we mainly focus on edges which are, potentially, the main culprits of high decoding failure rates. In particular, we define weights $w$ only for the edges that are present in most of the $4$-cycles in the Tanner graph of the code and by removing them the girth of the Tanner graph is increased to $6$. We call such edges ``culprit edges''. The weight for the non-culprit edges are considered to be $1$. So, the pair of weights to train for culprit edges is $(w,w')$ and for the other edges is set to be $(1,w')$. By properly training these weights and running SPA with the trained weights, we achieve an improvement in the decoding process of lattices constructed based on Construction A, whose underlying code is a HDPC code.
	
	The rest of the paper is organized as follows. In Section \ref{II}, we give some basic notations and definitions. In Section \ref{III}, we present three contributions: first, the structure of the trellis graph, which is completely new and different from the ones in the literature; second, the message computations in variable node and check node operations on the trellis graph; and third, a few algorithms to train weights in the neural network decoding algorithm. In Section \ref{IV}, we apply our newly proposed deep neural network decoding algorithm to a lattice  based on Construction A. We finish our paper in the last section giving concluding remarks and open problems. 
	
	\section{Preliminaries}\label{II}
	A lattice, $\Lambda$, is a discrete additive subgroup of $\mathbb{R}^m$ \cite{Conway}. The set of linearly independent vectors $\{{\bf b}_1,{\bf b}_2,\dots,{\bf b}_n\}$, where $n$ is the dimension of the lattice, is a lattice basis. 
	The generator matrix of the lattice is defined as ${\bf B}=[{\bf b}_1,{\bf b}_2,\dots,{\bf b}_n]^T$ and a lattice can be shown by $\Lambda=\{{\bf v}={\bf x}{\bf B}:\ {\bf x}\in\mathbb{Z}^n\}$. We let $m=n$ and consider only full dimensional lattices in this paper. The length of the shortest nonzero vector of the lattice is denoted by $d_{\min}$ with respect to Euclidean norm. The notation $\det(\Lambda)$ is used to denote the volume of the lattice which is obtained by $\det({\bf B}{\bf B}^T)$, and $V_n$ is the volume of the $n$-dimensional sphere of radius 1. For an unconstrained additive with Gaussian noise (AWGN) channel, the volume-to-noise ratio of the lattice $\Lambda$ is defined by $\mathrm{VNR}=\frac{(\det(\Lambda))^{\frac{2}{n}}}{2\pi e\sigma^2}$. 
	
	Suppose $C\subseteq\mathbb{F}_p^n$ is a linear code over $\mathbb{F}_p^n$, where $p$ is a prime number. We construct $\Lambda$ as follows: ${\bf x}=(x_1,\dots,x_n)$ is a point of lattice if and only if there exists a codeword ${\bf c}=(c_1,\dots,c_n)$ such that ${\bf x}\equiv{\bf c}\bmod p$. In other words,
	\begin{equation}\label{ConstructionA}
	\Lambda\!=\!p\mathbb{Z}^n+C\!=\!\{(pz_1+c_1,\dots,pz_n+c_n)\colon z_i\in\mathbb{Z},{\bf c}\in C\},
	\end{equation}
	where the components of ${\bf c}$ in (\ref{ConstructionA}) are considered as real numbers and operations are done in reals. The produced lattice $\Lambda$ is said to be constructed based on Construction A \cite{Conway}. 
	

	\section{Trellis Graph and Back-Propagation Algorithm}\label{III}
	Every linear code with parity check matrix ${\bf H}$ can be represented by a Tanner graph which is a bipartite graph 
	whose incident matrix is ${\bf H}$, see Example 1. In the existing iterative decoding algorithms in the literature, a message passing process is considered on the Tanner graph. The SPA is a message passing algorithm that exchanges messages through the edges in the corresponding Tanner graph. In this section, we first explain SPA on Tanner graphs and trellis diagrams and then we introduce our back-propagation algorithm. 
	\subsection{Sum-Product Algorithm on Tanner Graph}
	Let us first recall a SPA on Tanner graph of a code. Each iteration in SPA consists of two main processes, variable node process and check node process. Let $v_i$, $ch_j$, $c_v$ and $y_v$ denote  the $i$-th variable node, the $j$-th check node, the $v$-the component of the transmitted codeword and the $v$-th channel output, respectively.
	Nodes $v_i$, $1\leq i\leq n$, are associated with the initial inputs which are the log-likelihood ratios (LLR) of the channel output given by   
	\begin{equation}
	\ell_v=\log\frac{\mathrm{Pr}(c_v=1|y_v)}{\mathrm{Pr}(c_v=0|y_v)}.
	\end{equation} 
	
	In the variable node operation, the message $\mu_{v_i,ch_j}$ from $i$-th variable node to $j$-th check node is computed by 
	\begin{equation}\label{VNOp}
	\mu_{v_i,ch_j}=\ell_v+\sum_{
		ch_k\in V_i,\ k\neq j
	}\mu_{ch_k,v_i},
	\end{equation}
	where $V_i$ is the set of all neighboring check nodes to the variable node $v_i$. In the check node operation, the message $\mu_{ch_j,v_i}$ from the $j$-th check node to the $v_i$ is computed by:
	\begin{equation}\label{CNOp}
	\mu_{ch_j,v_i}=2\tanh^{-1}\left(\prod_{v_k\in CH_j,\ k\neq i}\tanh\left(\frac{\mu_{v_k,ch_j}}{2}\right)\right),
	\end{equation}
	where $CH_j$ denotes the set of all neighboring variable nodes to the $j$-th check node $ch_j$.
		After each check node operation we have a hard decision process in which  we obtain updated LLRs, denoted by $o_v$, for $1\leq v\leq n$, and computed as follows:
	\begin{equation}\label{HD}
	o_v=\ell_v+\sum_{
		ch_k\in V_i
	}\mu_{ch_k,v_i}.
	\end{equation}
	
	If $o_v<0$, then $c_v=0$, and if $o_v>0$, then $c_v=1$. If ${\bf H}{\bf c}^T=0$, then the decoding process halts and returns the decoded codeword.  
	\subsection{Sum-Product Algorithm on Trellis Graph}
	In the neural network decoding algorithms proposed in \cite{Learning2016}  and \cite{Learning1}, the message passing process is taken place on the trellis graph corresponding to the Tanner graph. The number of hidden layers with $L$ full iterations is $2L$; with the input and output layers there are a total of $2L+2$ layers in the trellis graph. In the following, we introduce a new trellis graph representation, which is different from the ones in the literature.
	
	Let $n$ and $E$ be the number of variable nodes and the number of edges in the Tanner graph of a linear code, respectively. If $L$ is the number of iterations, then the trellis graph has $3L+ 1$ layers. The first layer consists of $n$ nodes. There are three hidden layers for the $\ell$-th iteration. The first layer is $VN_\ell$, the second layer is $CN_\ell$, each of which contains $E$ nodes. The first and the second layers are associated with variable node and check node operations, respectively. The third layer is $O_\ell$ consisting of $n$ nodes. This layer is associated with the hard decision process.
	
	Before drawing edges between nodes in the trellis graph, we have to find culprit edges whose weights, $w$'s, should be trained. We explain this further in an example later. For each edge in the Tanner graph between variable node $v_i$ and check node $ch_j$, we define two types of edges in the trellis graph, $e_{v_i,ch_j}$ and $e_{ch_j,v_i}$, respectively, for $VN_\ell$ and $CN_\ell$, where $1\leq \ell\leq L$. We draw edges in the trellis graph using the following steps:
	\begin{itemize}
		\item If $ch_j\in V_i$, then  we connect node $e_{v_i,ch_j}$ of the layer $VN_1$ and the $i$-th node of the first layer.
		\item Each node $e_{ch_j,v_i}$ of the layer $CN_\ell$ in the $\ell$-th iteration is connected to $e_{v_k,ch_j}$ of the layer $VN_\ell$, where $v_k\in CH_j$ and $k\neq i$. For each node, we define a weight $w_{ch_j,v_i}$. 
		\item Each node $e_{v_i,ch_j}$ of the layer $VN_\ell$ in the $\ell$-th iteration is connected to $e_{ch_k,v_i}$ of the layer $CN_{\ell-1}$, where $ch_k\in V_i$ and $k\neq j$.  
		\item Each node $e_{ch_j,v_i}$ in the layer $CN_\ell$ is connected to $v_i$ in the layer $O_\ell$, where $ch_j\in V_i$. There is a weight, $w'_{ch_j,v_i}$, between the node $e_{ch_j,v_i}$ in the layer $CN_\ell$ and its corresponding variable node, $v_i$, in the layer $O_\ell$.
	\end{itemize}

	 Hence, we assigned a pair of weights $(w_{ch_j,v_i}$, $w'_{ch_j,v_i})$ to the edge $e_{ch_j,v_i}$; see the following example.
	\begin{Example}\label{Example2}
		Let $C$ be a linear code with parity-check matrix
		{\small \begin{equation}
		{\bf H}=\left[\begin{array}{ccc}
		1&1&1\\
		0&1&1\\
		\end{array}\right].
		\end{equation}}
		\begin{sloppypar}
		As we see in  Fig. \ref{Tanner}, there are five edges in its Tanner graph and four of them are involved in a unique $4$-cycle in the graph. We aim to train the weight $w$ for one of these four edges. Suppose the edge corresponding to $h_{22}$ is chosen  as a special edge. Hence, the pairs of weights which we consider to train are $(1,w'_{ch_1,v_1})$, $(1,w'_{ch_1,v_2})$, $(1,w'_{ch_1,v_3})$, $(w_{ch_2,v_2},w'_{ch_2,v_2})$, and $(1,w'_{ch_2,v_3})$ for the corresponding edges $e_{ch_1,v_1}$, $e_{ch_1,v_2}$, $e_{ch_1,v_3}$, $e_{ch_2,v_2}$, and $e_{ch_2,v_3}$, respectively. The corresponding  trellis graph with $2$ iterations is shown in Fig. \ref{Trellis}. 
		\end{sloppypar}
		\begin{center}
			\begin{figure}
				\centering
				\includegraphics[scale=.13]{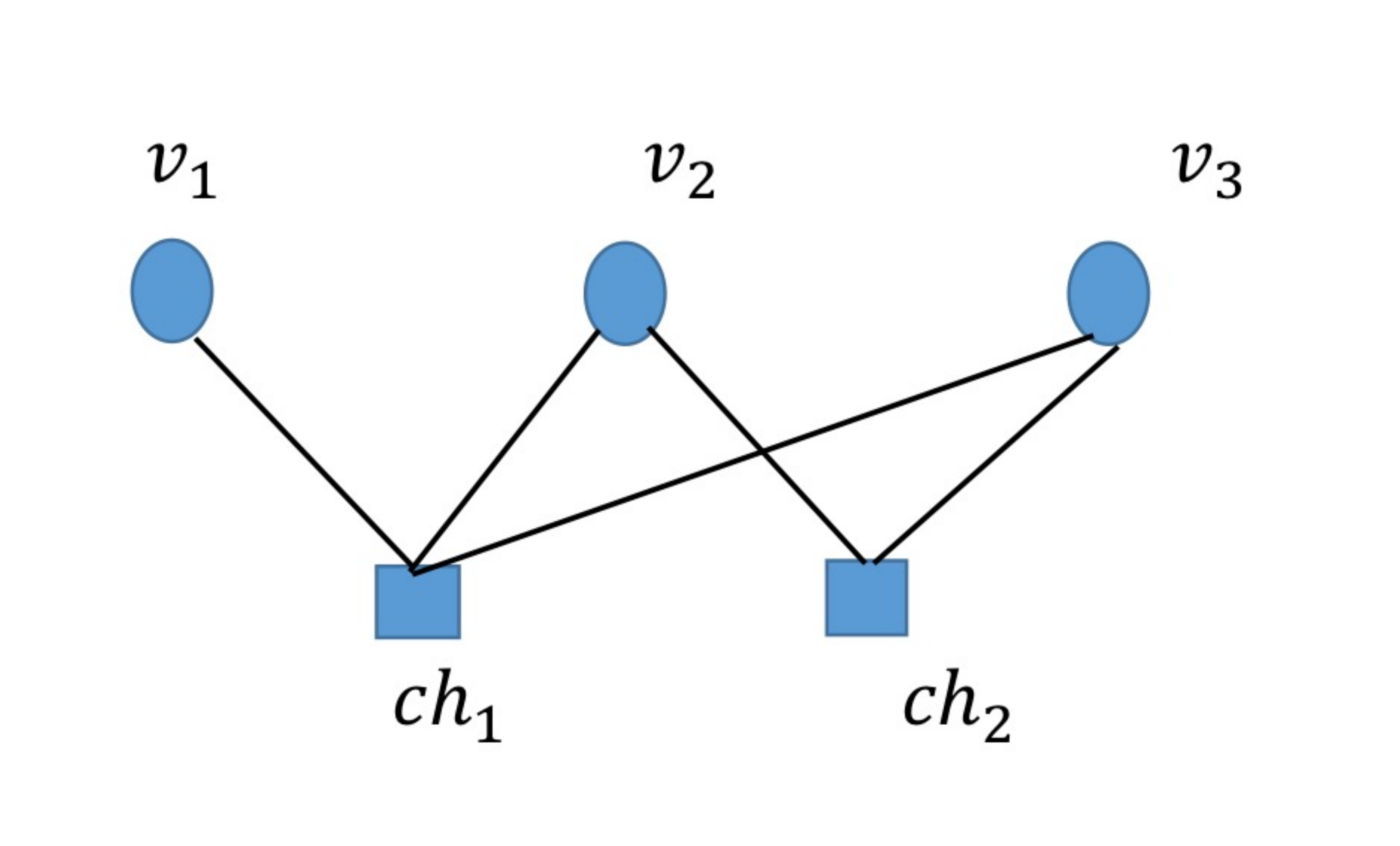}
				\caption{The Tanner graph of ${\bf H}$ in Example \ref{Example2}.}\label{Tanner}
			\end{figure}
		\end{center}
		\begin{center}
			\begin{figure}
				\centering
				\includegraphics[scale=.17]{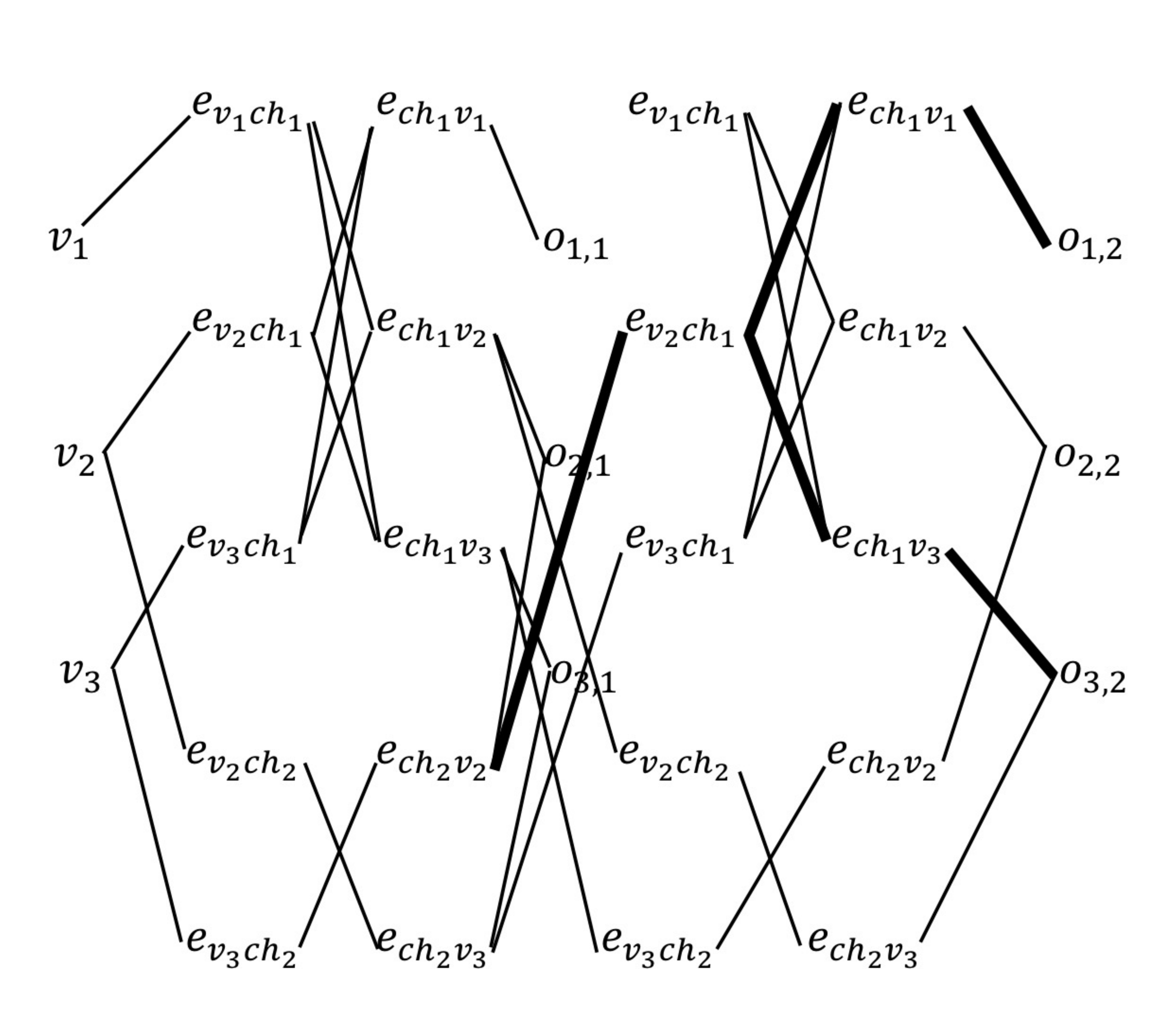}
				\caption{The trellis graph of ${\bf H}$ in Example \ref{Example2} with two iterations.}\label{Trellis}
			\end{figure}
		\end{center}
	\end{Example}
	\subsection{Back-propagation Algorithm}
	The back-propagation algorithm looks for the minimum of the error function in the weight space using the gradient descend method \cite{Rajosh}. The combination of weights, which minimizes the error function is the solution of the learning problem. This method requires computation of the derivative of the error function with respect to weights in each iteration step. Therefore, we have to guarantee the continuity and differentiability of the error function. The error function that we use is the cross entropy, which is defined as:
	\begin{equation}
	\mathcal{L}({\bf o},{\bf c})=\frac{-1}{n}\sum_{v=1}^{n}c_v\log_2(o_v)+(1-c_v)\log_2(1-o_v),
	\end{equation}
	where ${\bf o} = (o_1,\ldots,o_n)$, ${\bf c} = (c_1,\ldots,c_n)$, and $o_v$ and $c_v$ are $v$-th component of the deep neural network output and the transmitted codeword, respectively. If the transmitted codeword is ${\bf 0}$, then $c_v=0$ for all $v$.
	
	Moreover, in the back-propagation neural network, a sigmoid function is extensively used because it reduces complications involved during the training phase. The sigmoid function $\sigma$ is a real function from $\mathbb{R}$ to $(0,1)$ defined as
	$\sigma(x)=\frac{1}{1+e^{-x}}$.
	In the deep neural network decoder, each node of the hidden layers in the trellis graph contains two messages which we label them as the left and right messages. In the following, we explain how to obtain these messages for each node of the three hidden layers in the $\ell$-th iteration.
	\begin{itemize}
		\item  In $VN_\ell$ layer, the right message $\mu_{v_i,ch_j,\ell}$ of each node is defined as:
		\begin{equation}
		\tanh\left(\left(\ell_v+\sum_{ch_k\in V_i,\ k\neq j}w_{ch_k,v_i}\mu_{ch_k,v_i,(\ell-1)}\right)/2\right).
        \end{equation}
		The left message of each node is a vector of derivatives of $\mu_{v_i,ch_j,\ell}$ with respect to $\mu_{ch_k,v_i,(\ell-1)}$, $\frac{\partial \mu_{v_i,ch_j,\ell}}{\partial \mu_{ch_k,v_i,(\ell-1)}}$.
		\item 	In the $CN_\ell$ layer, the right message $\mu_{ch_j,v_i,\ell}$ of each node is  
		\begin{equation}
		2\tanh^{-1}\left(\prod_{v_k\in CH_j,\ k\neq i}\mu_{v_k,ch_j,\ell}\right).
		\end{equation}
		The left message of each node is a vector of derivatives of $\mu_{ch_j,v_i,\ell}$ with respect to $\mu_{v_k,ch_j,\ell}$, $\frac{\partial \mu_{ch_j,v_i,\ell}}{\partial \mu_{v_k,ch_j,\ell}}$.
		\item In the $O_l$ layer, the right message $o_{v,\ell}$ of each node is 
		\begin{equation}
		\sigma\left(\ell_v+\sum_{ch_k\in V_i} w'_{ch_k,v_i}\mu_{ch_k,v_i,\ell}\right).
		\end{equation}
		The left message of each node is a vector of derivatives of $o_{v,\ell}$ with respect to $\mu_{ch_k,v_i,\ell}$, $\frac{\partial o_{v,\ell}}{\partial \mu_{ch_k,v_i,\ell}}$.
	\end{itemize}

	We call the above constructed trellis graph whose nodes contain messages mentioned above a \emph{feed-forward neural network}. In fact, such a network represents a chain of function compositions, which transform an input to an output vector. The learning problem consists of finding the optimal combination of weights so that the decoded vector is as close as possible to the transmitted codeword. In other words, we aim to obtain weights by which the following error function $\mathcal{L}$ is minimized:
	\begin{equation*}
	\mathcal{L} = \sum_{\ell=1}^{L}\mathcal{L}({\bf o}_\ell,{\bf c}),
	\end{equation*}
	where ${\bf o}_\ell$ is the output of neural network at $\ell$-th iteration, $1\leq \ell \leq L$.
	Suppose the vector of weights are $W=(w_1,\dots,w_{E'},w'_1,\dots,w'_{E})$, where $E'<E$ is the number of culprit edges which are chosen to train their weights. We can minimize $\mathcal{L}$ by using an iterative process of gradient descent for which we need to calculate the gradient $\nabla \mathcal{L}$ as follows 
	\begin{equation*}
	\left(\frac{\partial \mathcal{L}}{\partial w_1},\dots,\frac{\partial \mathcal{L}}{\partial w_{E'}},\frac{\partial \mathcal{L}}{\partial w'_1},\dots,\frac{\partial \mathcal{L}}{\partial w'_{E}}\right).
	\end{equation*}
	
\begin{center}
 	\begin{figure}
 		\centering
 		\includegraphics[scale=.5]{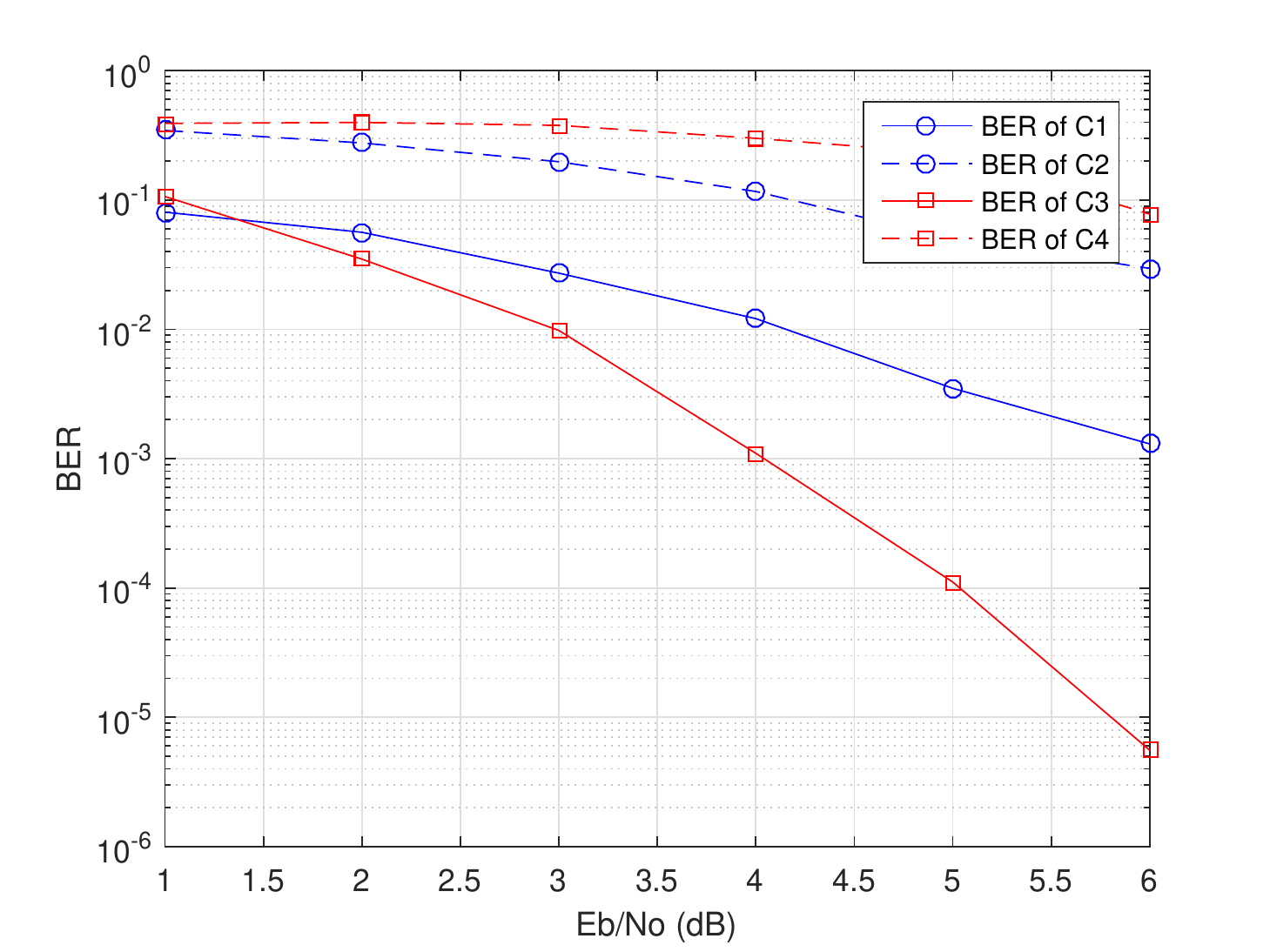}
 		\caption{The comparisons of performances of  $C1$, $C3$ and $C2$, $C4$.} \label{FIG4}
 	\end{figure}
 \end{center}

	If $\nabla \mathcal{L}\neq 0$, then we have to update weights and begin the process with new weights. We expect to find a minimum of the error function, when $\nabla \mathcal{L}= {\bf 0}$. In order to update weights, a learning constant (rate), denoted by $\alpha$, is used in the gradient descent algorithm. The learning rate is a proportionality parameter, which defines the step length of each iteration in the negative gradient direction.
	$W_{\mathrm{new}}=W_{\mathrm{old}}-\alpha\cdot\nabla \mathcal{L}$. 
	In the following example we illustrate how to update weights on the trellis graph.
	\begin{Example}
		Consider the parity-check matrix of Example \ref{Example2} with $L=2$ and the initial input ${\bf y}=(-0.5,2.5,-4)$ (received vector corresponding to sent codeword ${\bf c}={\bf 0}$) and with the initial weight vector $W=(0.1,0.15,0.16,0.17,0.18,0.19)$ in which the first element is associated to $w_{ch_2,v_2}$ and the other elements are associated to $w'_{ch_j,v_i}$. We also let the training rate be $\alpha=0.1$. We compute all messages both left and right in the trellis graph. Then, using the paths between the nodes of the last layer and nodes in the $CN_1$ and $CN_2$ layers in the trellis graph we update the weights. For example to update $w_{ch_2,v_2}$, we have to find paths from the last layer to the node $e_{ch_2,v_2}$ of the first iteration. As can be seen in Fig. \ref{Trellis}, we have two paths to this node, which are highlighted bold. 
		The paths $P_1$ and $P_2$ are $o_{1,2}\to e_{ch_1,v_1}\to e_{v_2ch_1}\to e_{ch_2,v_2}$ and $\ o_{3,2}\to e_{ch_1,v_3}\to e_{v_2ch_1}\to e_{ch_2,v_2}$, respectively. Therefore, the derivative of the function $\mathcal{L}$ with respect to $w_{ch_2,v_2}$ is $\frac{\partial \mathcal{L}}{\partial w_{ch_2,v_2}}=-0.013967$.
		 Hence, the new weight for the edge $e_{ch_2,v_2}$ is $(w_{ch_2,v_2})_{\mathrm{new}}=(w_{ch_2,v_2})_{\mathrm{old}}-\alpha\times\frac{\partial \mathcal{L}}{\partial w_{ch_2,v_2}}=0.1-0.1\times(-0.013967)=0.101396$. Continuing this process we obtain the new weights $W$ as:
	\begin{equation*}
	    (0.101396, 0.099499,  0.182551,  0.169208, 0.39523,  0.185566).
	\end{equation*}
	\end{Example}
Comparison between the works of \cite{Learning2016} and \cite{Learning1} shows that the former does not perform hard decision till the end of the last iteration 
while the latter performs hard decision in every iteration. Whereas, when it comes to decoding using SPA we perform hard decision in 
every iteration. However, we observed that in each variable node 
operation if instead of adding the  initial input to the 
incoming messages we add the updated LLRs from the previous 
iteration to the incoming messages, we obtain better performances. To show this, we compare the performance curves of two LDPC codes of sizes $10\times15$ and $63\times105$ using initial LLRs, $C2$, $C4$, respectively, and using updated LLRs, $C1$, $C3$, respectively. The difference between these curves is noticeable; see Fig. \ref{FIG4}.

	\section{Deep learning method to decode lattices constructed based on Construction A} \label{IV}
	The belief propagation algorithm obtains poor performance results compared to maximum likelihood algorithms if applied to HDPC codes. Therefore, to achieve better performances, we provide a deep neural network decoder for such lattices constructed based on Construction A. Using the same notation  proposed in \cite{Khodaie} and the deep neural network decoder presented in Section \ref{III}, we present the lattice decoding algorithm.  
	
	We take the volume-to-noise-ratio as $\mathrm{VNR}=\frac{4^{(2n-k)/n}}{2\pi e\sigma^2}$.  
	Suppose we send the all-zero codeword ${\bf c}$. We employ BPSK modulation to convert ${\bf c}$ into $(-1,\dots,-1)$. If the noise ${\bf n}$ is distributed as a normal distribution  $\mathcal{N}(0,\sigma^2)$ with zero mean and variance $\sigma^2$, then  the transmitted vectors are of the form ${\bf y}={\bf c}+4{\bf z}+{\bf n}$, \cite{Khodaie}. At the first step, ${\bf z}$ has to be decoded. We define the decoded ${\bf z}$ as $\hat{{\bf z}}=\lfloor \frac{{\bf y}-(1,\dots,1)}{4}\rceil$ as an all-one vector is added during encoding~\cite{Khodaie}. Now we define $a_i=y_i-4\hat{z}_i$, for $1\leq i\leq n$, and put $\hat{a}_i=2-a_i$ for $a_i>1$ and $a_i$ itself otherwise. Similar to \cite{Khodaie}, we need to define the $i$-th  input LLRs as	
	\begin{equation}
	\ell_i=\frac{(\hat{a}_i+1)^2}{2\sigma^2}-\frac{(\hat{a}_i-1)^2}{2\sigma^2}.
	\end{equation}
	Considering these LLRs as the received word of the deep neural network decoder, we decode ${\bf c}$ with  $\tilde{{\bf c}}$ as its decoded vector. We convert $\tilde{{\bf c}}$ to $\pm 1$ notation and call the obtained vector $\tilde{{\bf c}}'$. Finally, we let $\hat{c}_i=2-\tilde{c}_i'$ if $a_i>1$ and $\tilde{c}_i'$ otherwise.
	Then, the decoded lattice vector is $\hat{x}=\hat{c}+\hat{z}$. The stopping criterion in the back-propagation algorithm is $\nabla \mathcal{L}= {\bf 0}$, in other words we can get the trained weights when the trend of cross entropy changes from downwards to upwards. In practice, meeting this criterion is computationally costly. Instead, we impose a second termination condition. We define a parameter $\beta$ for the value of cross entropy which depends on $\mathrm{VNR}$. Therefore, the stopping criteria in our neural network decoding algorithm for lattices are either: 1) a change in the trend of the cross entropy, and/or 2) the value of cross entropy becomes smaller than $\beta$. Therefore, the two parameters $\alpha$ and $\beta$ determine the speed and precision of the experiment. The smaller parameters provide a better chance to train weights although, in this case, we need more time to get results. The learning rates chosen in \cite{Learning2016} and \cite{Learning1} are $\alpha=0.001$ and $\alpha=0.003$, respectively. 	
	
	\begin{Example}
		We applied our proposed neural network decoding algorithm with four iterations to decode Barnes-Wall lattice $\mathrm{BW}_8$ (see~\cite{Barnes}). We first determined the culprit edges in the Tanner graph of the Reed-Muller code, which is used in construction of $\mathrm{BW}_8$. The boldface $1$'s correspond to culprit edges. We train these edges using back-propagation algorithm to find the best possible weights. For initial weights, we used a vector of  length $33$ with normal distribution. We chose $\alpha=0.1$ and  $\beta=0.01$ for $\mathrm{VNR}=1$. By assuming weight $1$ for each edge, the bit error rate (BER) is 0.13 and the trained weights cause an improvement in BER, 0.098. For $\mathrm{VNR}=3$ and $6$, we let $\alpha=0.1$ and $\beta=0.001$ and $5\times(10)^{-5}$, respectively. An error performance comparison for $\mathrm{BW}_8$ between SPA in \cite{Khodaie} and our proposed decoding algorithm with four iterations is shown in Fig. \ref{FIG3}. It is clear that by reducing $\beta$ we can have better results for different $\mathrm{VNR}$'s.
		{\small \begin{equation}\label{Rela}
		{\bf H}=\left[\begin{array}{cccccccc}
		{\bf 1}&1&{\bf 1}&1&{\bf 1}&1&1&1\\
		{\bf 1}&0&1&0&1&0&1&0\\
		{\bf 1}&1&0&0&1&1&0&0\\
		{\bf 1}&1&1&{\bf 1}&0&0&0&0\\
		1&0&0&0&1&0&0&0\\
		1&0&1&0&0&0&0&0\\
		1&1&0&0&0&0&0&0
		\end{array}\right]
		\end{equation}}
\begin{center}
	\begin{figure}
		\centering
		\includegraphics[scale=.5]{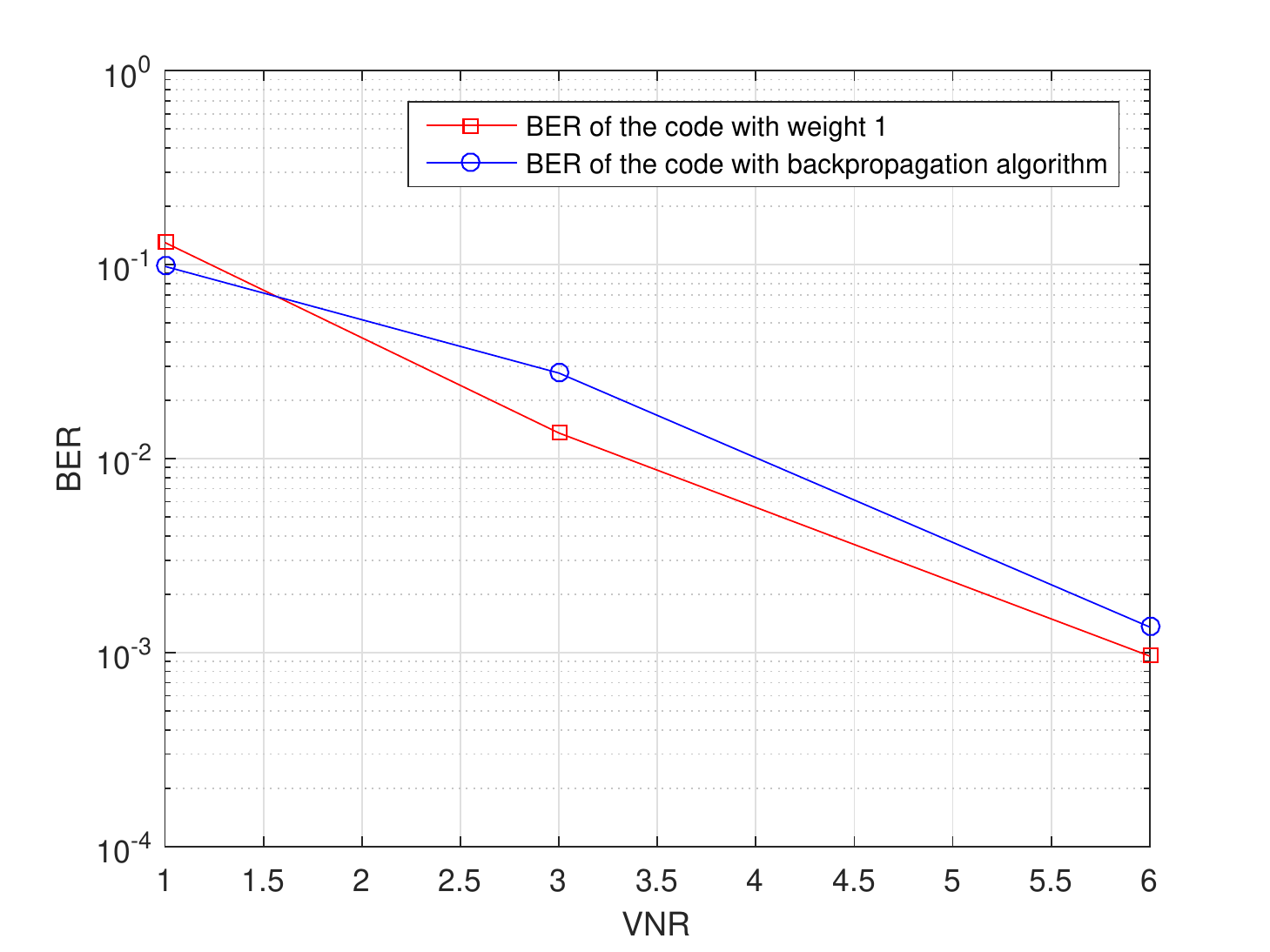}
		\caption{A comparison between the performance of SPA and back-propagation algorithm.} \label{FIG3}
	\end{figure}
\end{center}		
	\end{Example} 

The dependency of simulation results to the choices of parameters $\alpha$, $\beta$ and culprit edges is the topic of our future work. We also plan to extend neural network decoding algorithms to lattices constructed based on Construction D and D' with underlying HDPC codes such as Reed-Muller codes.

	\section{Conclusion}\label{V}
	In this paper, for the first time, a deep neural network decoding algorithm was presented for Construction A based lattices whose parity-check matrices are high-dense and have high decoding failure rates when iterative algorithms are used to decode them. In contrast with the neural network decoding algorithms in the literature, which were proposed for decoding linear codes and weights for all edges are required to be trained, we focused on training weights for edges which occur in most of 4-cycles and are, potentially, the main culprits of high decoding failure rates. Computer simulations were conducted to compare the error performance of the proposed algorithm versus a message passing algorithm.
	
\end{document}